\documentclass[12pt]{article}

\advance\hoffset by -1.1truecm
\advance\voffset by -1.0truecm

\def\be{\begin{equation}}
\def\ee{\end{equation}}
\def\ba{\begin{eqnarray}}
\def\ea{\end{eqnarray}}

\newcommand{\N}{\mbox{\rm N} \hspace{-0.9em} \mbox{\rm I}\,\,\,}
\newcommand{\R}{\mbox{I \hspace{-0.82em} R}}
\newcommand{\x}{{\bf x}}
\newcommand{\p}{{\bf p}}
\newcommand{\sn}{\smallskip\newline}
\newcommand{\mn}{\medskip\newline}
\newcommand{\bn}{\bigskip\newline}

\begin{document}
\title{On the Structure of Space-Time at the Planck 
Scale\thanks{Talk presented at the 36th School
 for Subnuclear Physics, \it From
 the Planck Scale to the
Hubble Radius, \rm Erice, Sicily, September 1998}}
\author{Achim Kempf\thanks{Affiliation at time of conference:
Dept. of Appl. Math. \& Theor. Physics, and Corpus Christi College,
Cambridge University, U.K.}\\
Institute for Fundamental Theory, Department of Physics\\
University of Florida, Gainesville, FL 32611, USA\\
{\small Email:  kempf@phys.ufl.edu}}

\date{}

\maketitle

\vskip-6.5truecm

\hskip11.7truecm
{\tt UFIFT-HEP-98-30} 

\hskip11.7truecm
{\tt hep-th/9810215}
\vskip6.1truecm

\begin{abstract}
The set of space-time short-distance structures which can be
described through linear operators is limited to a few basic
cases. These are continua, lattices and a further short-distance
structure which implies an ultraviolet cut-off.
Under certain conditions, these cut-off degrees 
of freedom can reappear as internal degrees    
of freedom. We review the current status of the classification
and present new conjectures. 
\end{abstract}
$$ $$
%{\centerline{{\bf Contents: }}}

%\begin{enumerate}
%\item Introduction
%\item Encoding  Space-Time Information using Linear Operators
%\item Classes of Short-Distance Structures
%\item Physical Origins
%\item Mathematical Origins 
%\item Gauge transformations
%\item Conclusions
%\end{enumerate}
%$$ $$ 

{\flushleft{ \bf 1~~ Introduction}}
\medskip\newline
The extrapolation of quantum theory and general relativity to the 
Planck scale is known to indicate a limit to the validity of the
conventional notion of locality. This is because 
test particles of sufficiently  high energy-momentum
to resolve a distance as small as 
a Planck length, about $10^{-35}m$, are predicted
to gravitationally curve and thereby to significantly disturb the very
space-time structure which they are meant to probe.
The unifying theory of quantum gravity is therefore expected to reveal a
nontrivial notion of locality at such small scales. For example,
Hawking \cite{hawking} and others have suggested
space-time to be foam-like at the Planck scale. More recent
suggestions are in terms of
strings and branes, see e.g. \cite{polchinski}, or also in terms of
 noncommutative, or `quantum'  
geometries, see e.g. \cite{nc}.
\mn
At least at present, however, the structure of
 space-time at the Planck scale cannot be 
probed directly by experiment.
In this paper we therefore ask whether a classification  of
the set of  all short-distance structures which space-time may
possibly have - under  some reasonable assumptions - can be achieved.
\newpage
\noindent
The two main messages in this paper are: 
\begin{itemize}
\item On the basis of relatively general assumptions,
a classification of the potential short-distance structures of space-time
can be achieved. 
\item One sub class of these short-distance 
structures yields a natural ultraviolet cutoff
which is such that the cut-off degrees of 
freedom reappear as internal degrees 
of freedom with unitary gauge groups.
\end{itemize}
%%%---------------------------------------- 
{\flushleft{ \bf 2~~ Encoding  Space-Time
 Information using Linear Operators}}
\medskip\newline 
Our aim is to try a classification of the 
short-distance structures that space-time may possibly have.
The basic assumption which we will make is 
that the fundamental theory of quantum gravity - whatever
this theory may be - encodes space-time 
information using operators $X_i$ which are
\it linear. \rm Since the $X_i$ should allow 
an interpretation in terms of `space-time information' 
we further assume that the formal expectation 
values of these operators are real:
\be
\langle \phi \vert X_i \vert \phi \rangle \in 
\R  ~~~~~\mbox{for all}~~\vert \phi\rangle \in D
\ee
Here, the vectors $\vert \phi \rangle$ run through a dense
domain $D$ of the $X_i$ in a complex Hilbert space $H$.
Technically, this is to say 
that we assume the $X_i$ to be symmetric operators.

We formally use 
the Dirac notation for these operators and 
vectors, and occasionally we will
formally also use terminology of nonrelativistic quantum mechanics. 
The use of this notation and terminology
 is of course merely for ease
of writing. 
Our aim is to cover an as large as possible
 set of candidates for a fundamental theory of
quantum gravity - 
in effect we aim at covering all or at least
 a large part of all
theories which are linear as quantum theories, 
i.e. which obey a linear superposition principle. 
Let us therefore keep in mind not to make any assumptions about
the actual physical interpretation 
of the $X_i$ in a fundamental theory of quantum gravity, 
nor to assume any particular physical interpretation of
the Hilbert space $H$ on which the $X_i$ act.

In this way, our approach is  general enough, 
for example, to cover the case of the 
matrix model for M-theory, where $N$-dimensional 
matrices $X_i$ are given the
interpretation that the eigenvalues stand 
for `space-time information' in the form of coordinates 
of $D0$-branes. The situation after quantization and 
taking $N = \infty$ will still be covered by our classification.
We will come back to this case in the last section.

The question arises of course, whether interesting 
conclusions can at all be drawn from 
assuming merely that a fundamental theory of 
quantum gravity encodes space-time information
using operators $X_i$ which are linear and symmetric.
To see that this is the case, let us recall that 
even linearity is far from being trivial; in particular, 
a linear map is not necessarily continuous.

Consider, for example,
 a matrix operator $X_{ij}$ acting on a sequence 
 of column vectors $v_j^{(n)}$.
Even if the all $v^{(n)}$ and their limit 
are in the Hilbert space of square summables, one finds
in general that:  
\be
\lim_{N\rightarrow\infty} ~\lim_{n\rightarrow \infty}~
\sum_{j=1}^{N} ~X_{ij} v_j^{(n)} ~~\neq~~
\lim_{n\rightarrow\infty} ~\lim_{N\rightarrow \infty}~
\sum_{j=1}^{N} ~X_{ij} v_j^{(n)} 
\ee
This is because 
the existence of one pair of limits does not imply 
the existence of the other pair 
of limits (and, for a generic matrix, even 
if all limits exist they may not commute).

One may be tempted to discuss such phenomena away, 
assuming that 
in practice one should always be able to approximate 
with finite dimensional
matrices. Recall, however, that the canonical commutation relation
$[\x,\p]=i1$ already provides an example for the 
necessity of infinite dimensional representations:
if $\x$ and $\p$ were
$n$-dimensional, the trace of the commutator on the 
LHS would vanish - while the trace of the RHS
 would be $i \cdot n$, which is growing with the dimension.
%%%----------------------------------------------
{\flushleft{ \bf 3~~ Classes of Short-Distance Structures }}
\medskip\newline 
A symmetric operator $X_i$ is an operator
 who's expectation values are real.
If an operator $X_i$ is symmetric, it may also be self-adjoint.
In this case it has a discrete or continuous spectrum.  
Therefore, self-adjoint $X_i$ can describe 
the two well-known short-distance structures
of lattices and continua, or, of course, 
mixtures of lattices and continua, which also includes fractals.

In addition to the two short-distance structures 
of lattices and continua, symmetric operators
can also describe a third short-distance structure,
which was named `fuzzy' in \cite{ak-bialo}. 
We will discuss the physical 
motivation for this terminology in the next 
section, and we will also identify two
sub classes among the operators of the fuzzy type. 
Mathematically, the fuzzy case is the case of
 operators $X_i$ which are
simple symmetric. By definition, simple symmetric
 operators are symmetric but not self-adjoint,
not even on any subspace. 

In order to gain intuition into why symmetric 
operators need not be self-adjoint,
consider again a matrix operator $X_{ij}$ on some 
dense domain $D$. If $X$ is symmetric,
i.e. if all its expectation values are real, then clearly
$X_{ij}=X_{ji}^*$. This, however, does not imply
self-adjointness, i.e. unique diagonalizability. 
Consider, for example, 
the eigenvalue equation
\be
X_{ij} v_j (\xi) = \xi v_i(\xi)
\ee
One may naively expect that two solutions $v(\xi), v(\xi^\prime)$ for
two different eigenvalues $\xi\neq \xi^\prime$ 
are orthogonal. If $X$ is self-adjoint, their orthogonality of course
follows from 
\be
(\langle v(\xi)\vert X) \vert v(\xi^\prime)\rangle =
\langle v(\xi)\vert (X \vert v(\xi^\prime)\rangle). 
\ee
However, writing out this equation in components it becomes clear that 
it is not a consequence of symmetry, because in general we may have
\be 
\lim_{N_2\rightarrow \infty} ~\lim_{N_1 \rightarrow \infty}~
\sum_{i=1}^{N_1} ~\sum_{j=1}^{N_2} v_i^*(\xi) X_{ij} v_j(\xi^\prime)~
\neq ~
\lim_{N_1\rightarrow \infty} ~\lim_{N_2 \rightarrow \infty}~
\sum_{i=1}^{N_1} ~\sum_{j=1}^{N_2} v_i^*(\xi) X_{ij} v_j(\xi^\prime) .
\ee
since the two limits need not commute for a merely
 symmetric $X$. In this case 
some of the $v(\xi)$ may not be orthogonal,
in which case they cannot all be contained in the domain $D$ of $X$.
\mn
For the precise classification of the basic cases we can use
the so-called deficiency indices
\be
r_\pm = \mbox{dim}\left[\left(\left( X 
\pm i1\right).D\right)^{\perp}\right]
\ee
which were introduced by v. Neumann.
For self-adjoint operators $X_i$, i.e. for operators
 which describe lattices and continua,
both indices vanish, $r_+=r_-=0$. For operators $X_i$ which
describe fuzzy short-distance structures, i.e.
for simple symmetric operators 
there is at least one nonzero index. 
Let us distinguish two sub classclasses of the fuzzy
 cases, by referring to
the cases  $r_+ = r_-\neq 0$ as being of the type fuzzy-A,
and by referring to the cases $r_+ \neq r_-$ as 
being of the type fuzzy-B. 

In the generic case, of course, symmetric operators $X_i$ can
be self-adjoint and simple symmetric on different subspaces, i.e.
generic symmetric operators are able to describe 
arbitrary mixtures of the basic cases, namely arbitrary mixtures of 
lattices, continua and fuzzy short-distance structures.

%%%----------------------------------------
{\flushleft{ \bf 4~~ Potential physical 
origins of operators of the type fuzzy-A}}  
\medskip\newline 
In the following, let us focus attention on 
the short-distance structures of the type fuzzy-A. These
are described by simple symmetric operators
 $X_i$ with equal deficiency indices. We will
for the moment also assume the indices to be 
finite: $0 \neq r := r_+=r_- \in \N$.
These operators can also be characterized by a physically more
intuitive criterion, which will motivate the terminology `fuzzy':

As will be proven in \cite{ak-sds}, an 
equivalent definition of operators of the type fuzzy-A 
is the following:
At each real expectation value there exists
 a finite lower bound to the formal spatial
uncertainty. 
\mn
To be precise, we are using the conventional 
definition of the uncertainty or standard deviation
for normalized $\vert \phi\rangle$:
\be
\Delta X_{\vert\phi\rangle} = \langle \phi \vert
 (X - \langle \phi \vert X \vert \phi \rangle)^2\vert \phi
\rangle^{1/2}
\ee
Then, $X$ is of the type fuzzy-A exactly iff there
 exists a positive function
\be 
\Delta X_{\mbox{\tiny min}}(\xi) > 0,
\ee
so that for each $\xi \in \R$, 
all normalized $ \vert \phi \rangle \in D$ with expectation
$\langle \phi \vert X \vert \phi \rangle = \xi$ obey
\be
(\Delta X)_{\vert \phi \rangle} \ge \Delta X_{\mbox{\tiny min}}(\xi) .
\ee
In naive quantum mechanical terminology, operators of 
the type fuzzy-A therefore
describe spaces
in which even with an ideal measurement apparatus 
the uncertainty or standard deviation in positions 
could not be made smaller than some 
finite lower bound $\Delta X_{\mbox{\tiny min}}(\xi)$.
 Since the lower bound is in general some
function of the expectation value $\xi$
the amount of `fuzzyness' can vary from place to place.
\mn
There are indications from general quantum 
gravity studies and from string theory which
point towards the fuzzy-A type of 
short-distance structure. Several studies 
 suggest that the uncertainty
relations effectively pick up correction terms, see e.g.
\cite{ucrs}. In the simplest case these are of the form
\be
\Delta x \Delta p \ge \frac{\hbar}{2}\left( 1 +
 \beta ~(\Delta p)^2 + ... \right) \label{ucr}
\ee
where $\beta>0$.
For a sufficiently small constant 
$\beta$, the correction term is negligible at 
present-day experimentally accessible
scales. At very small scales, the
correction term implies a crucial new feature, namely
that $\Delta x$ is now finitely bounded from below by  
\be
\Delta x_{\mbox{\tiny min}} ~  =  ~  \hbar~ \sqrt{\beta}
\ee
i.e. for all $\Delta x, \Delta p$ obeying
 (\ref{ucr}), there holds
$\Delta x \ge \Delta x_{\mbox{\tiny min}}$.
Choosing for $\beta$ the inverse square of the 
Planck momentum yields for 
$\Delta x_{\mbox{\tiny min}}$ the Planck length.
A string scale is obtained by relating $\beta$ to $\alpha^\prime$.
For reviews, see e.g. \cite{garay,witten}.

The functional analysis of operators leading to such
generalized uncertainty relations
was first studied in \cite{ak-ucr}.
It was pointed out in \cite{ak-bialo} that 
\it any \rm linear operator $X$ 
which obeys an
uncertainty relation that yields a lower 
bound $\Delta X_{\mbox{\tiny min}}>0$,
within any arbitrary theory,   
must be of the fuzzy type, i.e. simple symmetric. 
Let us add that, more precisely, any such operator is
of the type fuzzy-A, i.e. simple symmetric with
equal deficiency indices.
\mn
We remark here only in passing that in the 
case of short-distance structures
of the type fuzzy-B, there exist sequences of 
vectors in the physical domain such that
 $\Delta x$ converges to zero.
These short-distance structures are `fuzzy' in the sense 
that vectors of increasing localization 
around different expectation values then 
in general do not become orthogonal.
This will be proven and discussed in detail in \cite{ak-idf}.

%%%---------------------------------------
{\flushleft{ \bf 5~~ Potential mathematical 
origins of operators of the type fuzzy-A}}
\medskip\newline 
We will here not try to speculate in detail how 
operators $X_i$ of the type fuzzy-A may
mathematically arise from a fundamental theory 
of quantum gravity. We can, however,
address an important general point:

Operators, and in particular discontinuous operators, are
only fully defined if also their domain is specified.
Readers familiar with functional analysis will 
know that symmetric operators with equal 
deficiency indices possess domain extensions on 
which the resulting operators are 
self-adjoint. One may e.g. recall cases where 
self-adjoint extensions of 
differential operators 
correspond  to choices of  boundary conditions of
 some physical system.
Therefore, the important question  arises
in which ways, mathematically, a theory can \it 
intrinsically \rm  specify and 
fix the domain of its operators $X_i$ 
to be a domain on which the $X_i$ are simple 
symmetric, even if self-adjoint extensions exist.

Let us here discuss only the perhaps most
 obvious way
in which a theory may \it intrinsically \rm 
fix the domain of the $X_i$, namely
through kinematical and dynamical operator equations:
Requiring operator equations in a theory to hold
implies, in particular, that only a  domain which 
is common to all operators which appear
in the equations can be a physical domain.

As a simple example, consider the stringy uncertainty
 relation of above. The uncertainty
relation may ultimately arise
in a complicated way from the fundamental theory, but
 for the purposes of this argument,
let us here model the origin of the uncertainty relation 
through a simple correction term to the
 canonical commutation relation:
\be
[\x,\p] =i\hbar (1 +\beta \p^2) \label{ccr}
\ee
To see that (\ref{ccr}) yields (\ref{ucr}), 
recall that $\Delta A \Delta B \ge 1/2 \vert \langle
[ A,B] \rangle\vert$ for any pair of symmetric
 operators $A$ and $B$ on a joint 
domain with their commutator, 
and that $\langle \p^2\rangle = (\Delta p)^2 + 
\langle \p\rangle^2$.
\mn
In principle, kinematical equations such as (\ref{ccr})
could be part of the theory. An equation such
 as (\ref{ccr}) would then indeed
determine that on all physical domains the operator 
$\x$ is simple symmetric. This is
because the lower bound $\Delta x \ge \hbar \beta^{1/2}$
 from the uncertainty relation
(\ref{ucr}) holds on any domain on which $(\ref{ccr})$ holds.
In self-adjoint domain extensions of $X$, on the other hand, 
there necessarily exist vectors of arbitrarily small 
$\Delta x$, due to the diagonalizability.
Therefore, a theory which included (\ref{ccr}) would 
intrinsically ensure that 
the self-adjoint extensions are outside any physical 
domain - a physical domain 
being defined as a domain on which the theory's equations hold,
including, here, equation (\ref{ccr}).

Similarily, in the fundamental theory, domain 
specifications for the $X_i$ may arise, 
for example, from any kinematical or dynamical 
operator equations among the $X_i$, which
may be noncommutative, and with any other 
operators in the theory.
 \bn
We remark that the method of modeling generalized 
uncertainty relations kinematically, through
corrections to the canonical commutation relations,
 has been studied in some detail,
and it has been applied to both quantum mechanical 
and to quantum field theoretical examples.
 Among the main results are the following: Examples 
 have been given \cite{ak-osc}, which
demonstrate that fuzzy-A type geometries need not 
break external symmetries, as opposed e.g. to 
lattices. Further, there is a path integral 
formulation of quantum field theories in
fuzzy geometries \cite{ak-qft1,ak-gm-qft}.   
Within this approach, ultraviolet regularity 
on fuzzy-A type geometries has been shown to arise
by the following mechanism \cite{ak-goslar}: 

Ordinarily, in position space, ultraviolet divergencies
are known to originate in the ill-definedness of products
 of propagators and vertices which,
in the position representation, are distributions.
 Propagators such as $G(x,x^\prime)=
\langle x\vert (\p^2 +m^2)^{-1} \vert x^\prime 
\rangle$ are distributions because the formal
position eigenfields $\vert x\rangle$ are nonnormalizable. 
(For a description of the operators $\x_i,\p_i$
and their Hilbert space of fields in the path integral, 
see e.g. \cite{dewitt}.)
In the fuzzy-A case, the fields
$\vert x\rangle$ which are of maximal localization 
around the expectation value $x$, are
generalized coherent states. As such, they are generally
 normalizable. Thus, if these are
used to define field theories which are as local as 
possible on the given geometry, the
resulting Feynman rules are regular functions who's products
are well defined, which then implies ultraviolet regularity.

%%%---------------------------------------------
{\flushleft{ \bf 6~~ Gauge transformations}}
\medskip\newline 
Let us recall that we are discussing generic 
theories, not necessarily quantum field theories,
of which we assume only that they
encode `space-time information' using linear 
symmetric operators $X_i$. 
We found that in theories in which the $X_i$ are of the type
fuzzy-A there exists a finite lower bound 
$\Delta X_{\mbox{\tiny min}}(\xi)$ to the 
formal uncertainty $\Delta X$. 
A short-distance structure of this type clearly
 affects the very notion of 
locality. We are therefore led to consider the
 implications, for example, for the 
\it local \rm gauge principle.
But will it be possible
to deduce any information regarding gauge 
symmetries from such general assumptions?

Since the only concrete tools at hand are
 the operators $X_i$, let us make the ansatz
to define \it local \rm gauge transformations 
as the set of isometries (linear
operators which  preserve the
scalar product in Hilbert space) which
map a physical domain onto a physical domain, 
and which commute with the operators $X_i$:
\be
G := \left\{ \mbox{isometric} ~ u:  D^\prime 
\rightarrow D^{\prime\prime}, ~~
[u,X_i] = 0, ~i=1,2,...,n, \mbox{~~where~~} 
D^\prime,D^{\prime\prime} \subset D
  \right\} 
\label{lgt}
\ee
We will use the conventional terminology,
 but again, let us be
careful not to assume any particular physical interpretation
or r{\^o}le which these transformations may
 play in a fundamental theory.
\sn
With the definition (\ref{lgt}), we cover 
familiar cases such as local  gauge 
transformations of the
form $g = \exp( i \alpha_j(X)T_j)$ 
where the $T_j$ generate,
e.g., a local $U(5)$ on an ordinary 
continuous space with an isospinor index:
$g$ as an operator is clearly unitary and
 it commutes with the $X_i$.

The definition (\ref{lgt}) for local gauge transformations
is also general enough to be applicable to 
the case of operators $X_i$ of the fuzzy types.
This is because in (\ref{lgt})   
the localness of a gauge transformation $u$ is defined through the 
criterion that $u$ commutes with the $X_i$, which is
a criterion that does not require the $X_i$ to be diagonalizable.
\mn
Let us apply the definition (\ref{lgt}) to
the case of  short-distance structures of the type fuzzy-A. 
We saw in the fuzzy-A case that those Hilbert space vectors, 
or `degrees of freedom', which would describe
structures smaller than the scale of 
fuzzyness $\Delta X_{\mbox{\tiny min}}(\xi)$ 
are cut-off from the domain of the $X_i$.
However, the cut-off degrees of freedom will
 nevertheless play an important r{\^o}le: 
As we will see, mathematically, it is the 
self-adjoint extensions which describe
 the cut-off degrees of freedom - 
and the self-adjoint extensions will give 
rise to an isospinor structure, i.e. 
internal degrees of freedom, automatically 
with unitary gauge groups. 
\sn
The underlying reason for the re-appearance 
of those degrees of freedom is that
whereas the $X_i$ are discontinuous
operators which, therefore, do not see the entire Hilbert space,
 isometries $u$ are necessarily bounded and
continuous operators. Because of their continuity,
no part of the Hilbert space can be hidden from such operators.
\mn
Let us consider the example of a single 
operator $X$ which describes
a short-distance structure of type fuzzy-A.

As mentioned already, operators $X$ 
which are of type fuzzy-A always 
have self-adjoint extensions
in the Hilbert space, though outside the physical domain. 
Each extension has its own discrete
spectrum and together the spectra can be shown to cover all reals.
Thus, in extensions, arbitrarily sharp localization around
arbitrary position expectation values can be reached.
In this sense, the family of self-adjoint extensions contains 
the degrees of freedom beyond the cutoff scale.

Crucially, the self-adjoint extensions, 
though outside the physical domain, 
do appear  in the construction of the gauge transformations,
thereby bringing back the cut-off degrees of freedom as
internal degrees of freedom:

As will be shown in \cite{ak-idf}, \it all 
\rm unitaries can be expressed as functions of the 
self-adjoint extensions of $X$. Unitary functions
$u$ of a self-adjoint extension $X_e$ of $X$ 
commute with $X_e$ and
any isometric restriction of $u$ which maps
a physical domain onto a physical domain, 
therefore, commutes with $X$, thus yielding
a gauge transformation according to our definition (\ref{lgt}).

These include in fact `local' $U(r)$- gauge 
transformations, where 
$r$ is the deficiency index.  The necessary
isospinor structure emerges automatically!  
To this end, it will be proven in \cite{ak-idf} that 
for each real $\xi$ there are self-adjoint extensions of $X$
for which $\xi$ is an $r$-fold degenerate
eigenvalue. This implies that for the 
(non-symmetric) adjoint operator
$X^*$ each real $\xi\in \R$ is an $r$- fold
 degenerate eigenvalue with eigenvectors 
$\vert \xi,i\rangle$ where $i=1,...,r$.
Any vector $\vert \phi\rangle$ can be 
represented by an isospinor function 
$\phi_i(\xi) = \langle \xi,i\vert \phi\rangle$, 
which shows the appearance of
the isospinor structure. At large scales, the 
eigenvectors become orthogonal
i.e. $g$ is local in the conventional sense;
at small scales the variations of $g$ are 
restricted by the physical domain condition in 
(\ref{lgt}).
\mn 
The mechanism by which internal symmetries arise
 here has some intuitive similarities 
with the mechanism by which internal symmetries arise
in the Kaluza-Klein approach: there, in the simplest case, 
at each point in space-time a little circle is 
attached, in an extra dimension.
Here, say in the simplest case of deficiency 
indices $(1,1)$, at each point 
a little $S^1$ exists. However, this $S^1$ is 
not in an extra dimension.
Instead, this $S^1$ is `within' the point - a 
point now being a little patch of fuzzyness of size
$\Delta X_{\mbox{\tiny min}}$ \cite{ak-euro}. 

To see this, we note that 
the self-adjoint extensions of a simple symmetric
 operator with finite and
equal deficiency indices form themselves a 
representation of a $U(r)$, which implies that
each eigenvalue has an orbit under this $U(r)$. In the simplest
case of deficiency indices (1,1) it reduces to a $U(1)$-orbit.
Each eigenvalue's $U(1)$-orbit
is just small enough not to be resolvable in 
the presence of the fuzzy cutoff. 

\newpage
{\flushleft{ \bf 7~~ Conclusions}}
\medskip\newline 
We investigated the classification of all short-distance
 structures which are describable 
by operators which are linear and have real expectation 
values. We found that the generic
short-distance structure which these operators can 
describe is a mixture of the basic cases of
lattices, continua, and fuzzy spaces. 

As indicated in the beginning, we are covering the 
case of the matrix model for 
M-theory. Even after the matrix elements of the $X_i$ 
become operators through quantization, and
after taking $N=\infty$, the $X_i$ are linear
 and symmetric operators.
Thus, the $X_i$ then still fall into the discussed classification. 

We note that if those
$X_i$ are of a fuzzy type, this
would mean that their theory cannot be a straightforward limit of 
a sequence of theories based on finite dimensional matrices. 
This is because for finite dimensional matrices 
the deficiency indices always vanish since 
symmetry and self-adjointness coincide.  
Indeed, as we saw, an infinite dimensional matrix 
theory can have quantum numbers  - the 
discussed isospinor structure arising in the fuzzy-A case -
which are not present in any finite dimensional approximation.

If the $X_i$ are found to be of type fuzzy-A, 
then the eigenvalues which in finite dimensions stand
for $D0$-brane coordinates would assemble into 
$U(1)$-, or generally $U(r)$-group orbits 
which are just small enough not to be resolvable in the fuzzy geometry.
\mn
While this picture is valid for each individual  $X_i$, we need to recall
that  in our analysis of the individual $X_i$ we 
so-far held the other
coordinates $X_j$ fixed for $j\neq i$. We did
 so for ease of the analysis
since otherwise the deficiency indices would
 generically be infinite.
In general, of course, the short-distance 
structure may vary arbitrarily in an $n$-dimensional space
and, in particular, the $X_i$ may not commute. This
is reflected by the fact that the functional analysis
 of each operator $X_i$ is generically a function
of the functional analysis of the other $X_j$, 
as we discussed briefly
in section 5 in the context of mechanisms by which
theories can intrinsically specify operator domains.

The question of internal symmetries in the fuzzy-B case,
and  numerous further issues, such as
the interplay of fuzzy short-distance structures 
with supersymmetry and compactifications, remain to be addressed.

We remark that, so-far, classically real variables 
have been assumed to correspond to
self-adjoint operators also within the framework 
of noncommutative geometry. 
It should be very interesting to investigate the
 application of the tools of noncommutative geometry
to the general case of symmetric $X_i$.

Finally, we note that the fuzzy short-distance
 structure has recently been studied 
in the context of the transplanckian energy problem
 of black hole radiation \cite{broutetal}.

\end{document}